\documentstyle[epsf]{mn}

\title[]{First X-ray observations of the polar CE Gru}

\author[Ramsay \& Cropper]{
Gavin Ramsay and Mark Cropper\\
Mullard Space Science Lab, University College London,
Holmbury St. Mary, Dorking, Surrey, RH5 6NT, UK\\}

\date{}
\begin{document}
\outer\def\gtae {$\buildrel {\lower3pt\hbox{$>$}} \over 
{\lower2pt\hbox{$\sim$}} $}
\outer\def\ltae {$\buildrel {\lower3pt\hbox{$<$}} \over 
{\lower2pt\hbox{$\sim$}} $}
\newcommand{\ergscm} {ergs s$^{-1}$ cm$^{-2}$}
\newcommand{\ergss} {ergs s$^{-1}$}
\newcommand{\ergsd} {ergs s$^{-1}$ $d^{2}_{100}$}
\newcommand{\pcmsq} {cm$^{-2}$}
\newcommand{\ros} {\sl ROSAT}
\newcommand{\exo} {\sl EXOSAT}
\newcommand{\xmm} {\sl XMM-Newton}
\def\rchi{{${\chi}_{\nu}^{2}$}}
\newcommand{\Msun} {$M_{\odot}$}
\newcommand{\Mwd} {$M_{wd}$}
\def\Mdot{\hbox{$\dot M$}}
\def\mdot{\hbox{$\dot m$}}

\maketitle

\begin{abstract}

We report the detection of the polar CE Gru in X-rays for the first
time. We find evidence for a dip seen in the hard X-ray light curve
which we attribute to the accretion stream obscuring the accretion
region in the lower hemisphere of the white dwarf. The X-ray spectrum
can be fitted using only a shock model: there is no distinct soft
X-ray component. We suggest that this is because the reprocessed
component is cool enough so that it is shifted into the UV. We
determine a mass for the white dwarf of $\sim$1.0\Msun.

\end{abstract}


\section{Introduction}

CE Gru (also known as Grus V-1 and Hawkins V-1) was discovered by
Hawkins (1981, 1983) during a search for variable objects using UK
Schmidt plates. Further observations by Tuohy et al (1988) confirmed
its binary nature (an orbital period of 108.5 mins) while Cropper et
al (1990) found strong circular polarisation -- a characteristic of
the polar (or AM Her) class of cataclysmic variable. These objects are
interacting binaries in which material flows from a dwarf main
sequence star onto a magnetic ($B\sim$10--200MG) white dwarf. This
strong magnetic field is high enough to force the spin of the white
dwarf to be synchronised with the binary orbital period.

Tuohy et al (1988) demonstrated that there are two accretion poles
which are characterised by different emission properties: one pole
which is always in view and is stronger in blue light, while the other
pole which is visible for only $\sim$0.35 of the white dwarf spin
period is stronger in red light. Cropper et al (1990) showed that the
pole always in view was positively circularly polarised, with the
other being negatively polarised.

The UK Schmidt plates show that CE Gru exhibits two distinct levels of
brightness, ($B\sim$18 and $B\sim$21). In the fainter state the
accretion flow is much reduced, or stopped altogether. This may be the
reason why CE Gru was not detected during the {\ros} all-sky survey
(Verbunt et al 1997). Indeed, CE Gru has never before been detected in
X-rays. In this paper, we report the first X-ray detection of CE Gru
which were made as part of a survey of polars using {\xmm}.

\section{Observations}

The satellite {\xmm} was launched in Dec 1999 by the European Space
Agency. It has the largest effective area of any X-ray satellite and
also has a 30 cm optical/UV telescope (the Optical Monitor, OM: Mason
et al 2001) allowing simultaneous X-ray and optical/UV coverage. CE
Gru was observed using {\xmm} on 2001 Oct 31. The EPIC instruments
(imaging detectors covering the energy range 0.1--10keV with moderate
spectra resolution) were operated in full frame mode (the count rate
was not high enough to cause pile-up problems). The RGS detectors
(high resolution spectrographs operating in the 0.3--2.0keV range: den
Herder et al 2001) were configured in the standard spectroscopy
mode. We clearly detect CE Gru in the X-ray band. OM data were taken
in two UV filters (UVW1: 2400--3400 \AA, UVW2: 1800--2400 \AA) and one
optical band ($V$ band). The observation log is shown in Table
\ref{log}.

The data were processed using the {\sl XMM-Newton} {\sl Science
Analysis Software} v5.2. The RGS spectra were of low signal to noise
and showed no evidence for significant line emission: we do not
consider them further. For the EPIC pn detector (Str\"{u}der et al
2001), data were extracted using an aperture of 40$^{''}$ arc sec
centered on the source. Background data were extracted from a source
free region. For the EPIC MOS detectors (Turner et al 2001) we
extracted data in a similar way, but extracted the background from an
annulus around the source. The background data were scaled and
subtracted from the source data. In extracting the EPIC pn spectrum,
we used only single pixel events and used the response file
epn\_ff20\_sY9\_thin.rmf. In the case of the MOS data we used the
response files m[1-2]\_thin1v9q19t5r5\_all\_15.rsp. The OM data were
analysed in a similar way using {\tt omichain} and {\tt omfchain}
(this latter task was not incorporated in SAS v5.2 but will be in a
later version). Data were background subtracted and corrected for
coincidence losses (Mason et al 2001).

The Optical Monitor data shows that CE Gru had a mean brightness of
$V$=17.9 and a maximum $V$=17.5. Since the $V$ band observations
started at the descent from maximum its likely that its true maximum
was brighter than this. Assuming a similar colour to that found by
Tuohy et al (1988) ($B-V$=0.53) this places CE Gru in a high accretion
state at the time of the {\xmm} observations and a similar brightness
to observed by Tuohy et al ($V\sim$18). The length of the observation
in the EPIC MOS detectors covered just over 1 orbital cycle.

The flux in the UV filters corresponds to: UVW1
$\sim~4.5\times10^{-16}$ \ergscm \AA, UVW2 $\sim~4.0\times10^{-16}$
\ergscm \AA, (based on OM observations of isolated white dwarfs). The
mean flux in the $V$ filter corresponds to $\sim~2.5\times10^{-16}$
\ergscm \AA\ at 5000 \AA.

\begin{table}
\begin{center}
\begin{tabular}{lrrr}
\hline
Detector & Mode & Filter &Exp (s)\\
\hline
EPIC MOS & Full frame &thin& 7792 \\
EPIC pn & Full frame & thin & 5468 \\
RGS & Spectroscopy & & 8255\\
OM & Image/fast & UVW1 & 1500 \\
OM & Image/fast & V & 1500 \\
OM  & Image/fast & UVW2 & 1800 \\
\hline
\end{tabular}
\end{center}
\caption{The log of {\xmm} observations of CE Gru.}
\label{log}
\end{table}

\begin{figure}
\begin{center}
\setlength{\unitlength}{1cm}
\begin{picture}(8,10)
\put(0,-1.){\includegraphics{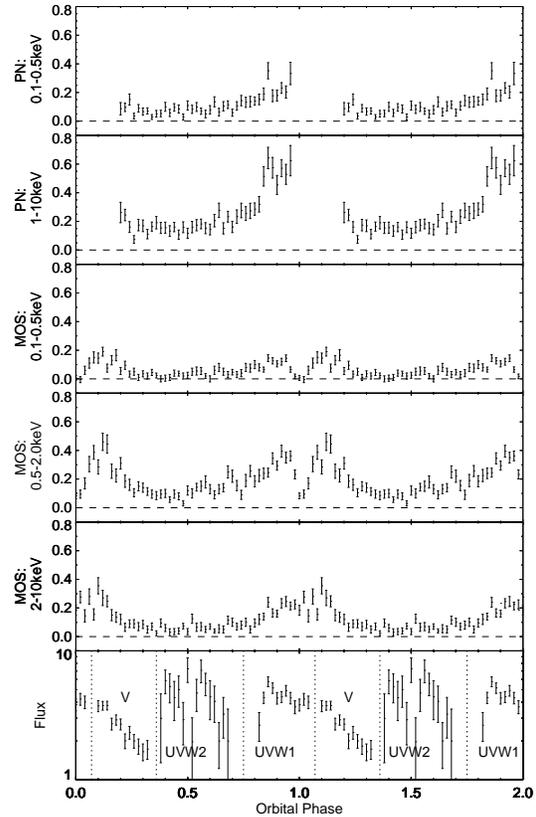}}
\end{picture}
\end{center}
\caption{The light curves obtained of CE Gru using {\xmm}. The X-ray
data were obtained using the EPIC pn and MOS detectors. The MOS1 and
MOS2 light curves have been co-added. The Optical Monitor data (bottom
plot) shows the optical/UV data with units of flux of $10^{-16}$
\ergscm. The data have been phased on the orbital period of Tuohy et
al (1988) and binned into $\delta\phi=$0.02 bins. We have chosen phase
0.0 to correspond to the dip centered in the bright phase.}
\label{light} 
\end{figure}

\section{Light curves}

We show in Figure \ref{light} the light curve of CE Gru in various
energy bands using EPIC pn and EPIC MOS data (where the MOS1 and MOS2
data have been co-added) folded on the orbital period of Tuohy et al
(1988). We also show the OM data: CE Gru was detected in all three
filters (it is also the first time that it has been detected in the
near-UV).

There are two distinct parts to the orbital light curve: a faint phase
lasting $\sim 0.6$ orbital cycles, and a brighter phase lasting $\sim
0.4$ orbital cycles. These relative durations are similar to those in
the optical light curves seen by Cropper et al (1990). The bright
phase is also of a similar duration to that of the red pole seen by
Tuohy et al (1988) when the system was at a similar brightness
compared to our observations. We therefore assign the bright phase
X-ray emission to the pole in the lower hemisphere, and the fainter
phase emission to the pole in the upper hemisphere. Although there
maybe some ambiguity in this assignment, the fact the we observe
decreasing emission in the $V$ band at the end of the bright phase
provides supporting evidence for this assignment since the `blue' pole
does not show such a rapid drop in $V$.

In the bright phase, a dip is seen in the soft X-ray light curve, but
not at higher energies. This is characteristic of photo-electric
absorption, and is seen in other polars (eg Watson et al 1989). It is
thought to result from the accretion stream crossing our line of sight
to the accretion region, thereby absorbing the soft X-rays. This is
consistent with the model of Wickramasinghe et al (1991) for CE Gru in
which the two accretion regions lie near the magnetic poles,
approximately at the foot-points of the field lines passing through
the region where the stream threads onto the magnetic field. It is
most likely that the dip is caused by the accretion stream to the {\it
upper} pole absorbing the emission from the lower pole: there is clear
evidence for accretion at the upper pole and it is inevitable that the
stream crosses our line of sight. It is possible for the ballistic
stream or the stream to the lower pole to be the cause of the
absorption, but this requires a high inclination and the ballistic
stream to penetrate very close to the white dwarf before threading.
These are special conditions, and in the absence of even a grazing
eclipse we consider this to be unlikely.

At energies greater than 2keV, there is no evidence for the dip seen
at lower energies. Using the spectral model described below we can
estimate the absorbing column density of the accretion stream which
causes the dip by increasing the column until we match the observed
count rate at lower energies. We estimate that the total column
density in our line of sight must be $\sim5-10\times10^{21}$
\pcmsq. This is of the same order as that seen in other polars with
stream dips (Watson et al 1989).

\section{X-ray spectra}

\subsection{The model}

We extracted a faint and bright phase spectrum ($\phi$=0.2--0.8 and
$\phi$=0.8--1.2 respectively) from the EPIC MOS data and a faint phase
spectrum from the EPIC pn data (these data covered too short a length
of time to obtain a useful bright phase spectrum). We modelled the
data using a simple neutral absorber and an emission model of the kind
described by Cropper et al (1999). This emission model, unlike single
temperature thermal bremsstrahlung models, is a realistic physical
description of the post-shock accretion region in polars. It is based
on the prescription of Aizu (1973) which predicts the temperature and
density profiles over the height of the accretion shock. However, it
has been modified to take into account cyclotron cooling (which can be
significant in polars) and also the variation in gravitational force
over the shock height. To reduce the number of free parameters we fix
the parameter, $\epsilon_{s}$, (the ratio of cyclotron cooling to
thermal bremsstrahlung cooling), at 5: this implies a magnetic field
strength typical of polars (20--50MG). We also fix the specific
accretion rate at 5 g s$^{-1}$ cm$^{-2}$ (typical of polars in a high
accretion state). Changing these parameters does not have a great
effect on the results for data with low-moderate signal to noise ratio
(as in our CE Gru data). Observations of polars using {\ros} showed
that many polars had prominent blackbody components (Ramsay et al
1994). Therefore, we also added a cool blackbody model and determined
the change in the fit.

\subsection{Results}

\begin{figure}
\begin{center}
\setlength{\unitlength}{1cm}
\begin{picture}(8,7)
\put(-1.,-0.5){\includegraphics{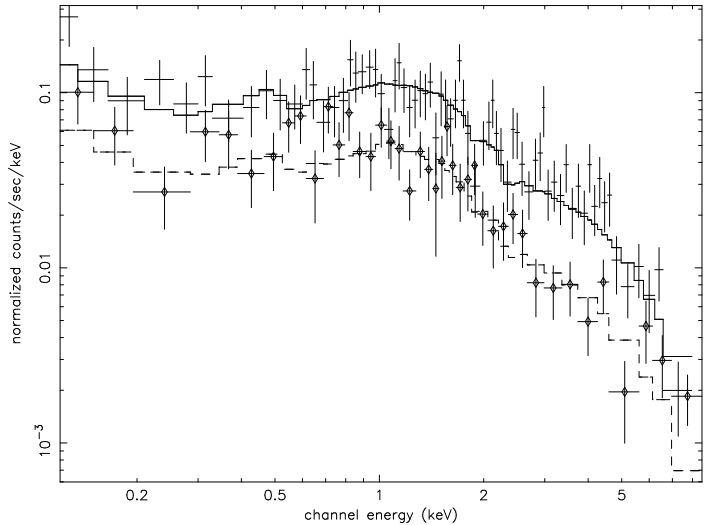}}
\end{picture}
\end{center}
\caption{The EPIC MOS1 spectra extracted from the bright phase (dots)
and the faint phase (diamonds) data together with a multi-temperature
shock model (solid line for the bright phase and dashed line for the
faint phase).}
\label{spec} 
\end{figure}

We initially consider the EPIC spectra from the faint state. We show
in Table \ref{specfit} the fits to the MOS and pn spectra (we fit the
MOS spectra simultaneously while leaving their normalisation
parameters untied). We find that the spectrum can be well modelled
without a blackbody component. The Hydrogen column density is low
($<10^{20}$ \pcmsq). Adding a blackbody to the model makes no
significant difference to the fit. We do, however, show in Table
\ref{specfit} the fit and flux when we add a blackbody of temperature
30eV (typical of the temperature of the blackbody component seen in
{\ros} observations of polars) to the model. In the bright phase data,
again we find no significant improvement to the fit when we add a
blackbody to the model. We show in Figure \ref{spec} the EPIC MOS1
spectra taken from the bright and faint phase, together with the best
fits (assuming no blackbody).

Table \ref{specfit} shows that the inferred mass of the white dwarf
(assuming the white dwarf mass-radius relationship of Nauenberg 1972)
is $\sim$1.0\Msun. Although this is higher than found for isolated
white dwarfs, it is comparable with the mass of other accreting
magnetic white dwarfs. Ramsay (2000) show that the latter class are
biased towards higher masses.

\begin{table*}
\begin{tabular}{lrrr}
\hline
  & \multicolumn{2}{c}{Faint Phase}& Bright Phase\\
  & EPIC MOS & EPIC pn & EPIC MOS\\
\hline
$N_{H}$ (10$^{20}$ \pcmsq) & 0.0$^{+1.3}$ & 0.0$^{+0.04}$ & 0.0$^{+0.02}$\\
$M_{1}$ (\Msun) & 1.04$^{+0.18}_{-0.23}$ & 1.03$^{+0.18}_{-0.15}$ & 
1.29$_{-0.15}$\\
Flux$_{hard,bol}$ (\ergscm) & 2.1$^{+0.9}_{-0.6}\times10^{-12}$ &
 2.2$^{+0.5}_{-0.3}\times^{-12}$ & 7.1$^{+0.8}_{-1.0}\times10^{-12}$\\ 
 & 2.1$^{+1.0}_{-0.6}\times10^{-12}$ & & 8.2$_{-1.4}^{+0.2}\times10^{-12}$\\
$L_{hard,bol}$ (\ergsd) & 2.5$^{+1.1}_{-0.7}\times10^{30}$ & 
2.7$^{+0.5}_{-0.4}\times10^{30}$  & 9.1$^{+0.2}_{-1.8}\times10^{30}$\\
 & 2.6$^{+1.2}_{-0.7}\times10^{30}$ & & 1.0$^{+0.1}_{-0.2}\times10^{31}$\\
\rchi (dof) & 1.07 (82) & 1.08 (104) & 0.92 (143)\\
+ bb (eV) & 30 & 30 & 30\\
Flux$_{soft,bol}$ (\ergscm) & 1.5$^{+1.2}_{-1.5}\times10^{-13}$ &
  1.9$^{+17}_{-1.9}\times10^{-13}$ & 1.1$^{+2.0}_{-1.1}\times10^{-12}$\\
 & 1.5$^{+1.2}_{-1.5}\times10^{-13}$ & & 8.9$^{+14.6}_{-8.9}\times10^{-13}$\\ 
$L_{soft,bol}$ (\ergsd) & 9.0$^{+1.8}_{-9.0}\times10^{28}$ &
  1.1$^{+1.7}_{-1.1}\times10^{29}$ & 1.9$^{+3.5}_{-1.9}\times10^{30}$\\
 & 9.0$^{+1.8}_{-9.0}\times10^{28}$ & & 1.6$^{+2.5}_{-1.6}\times10^{30}$\\
 \rchi (dof) & 1.06 (79) & 1.07 (103) & 0.90 (140)\\
\hline
\end{tabular}
\caption{The parameters for the EPIC spectra extracted from the faint
phase and bright phase. The MOS1 and MOS2 spectra were fitted
simultaneously with all the parameters linked except the
normalisations. We also show the effect on the fit if we add a
blackbody with $kT_{bb}$=30eV to the model.}
\label{specfit}
\end{table*}

\subsection{Soft and hard luminosities}

We define the hard X-ray luminosity as
($L_{hard}=4\pi$Flux$_{hard,bol}d^{2}$) where Flux$_{hard,bol}$ is the
unabsorbed, bolometric flux from the hard component and $d$ is the
distance. Since a fraction of this flux is directed towards the
observer, we switch the reflected component to zero after the final
fit to determine the intrinsic flux from the optically thin post-shock
region.  We define the soft X-ray luminosity as
($L_{soft}=\pi$Flux$_{soft,bol}$sec($\theta)d^{2}$), where we assume
that the soft X-ray emission is optically thick and can be
approximated by a small thin slab of material, the unabsorbed
bolometric flux is Flux$_{soft,bol}$ and $\theta$ is the mean viewing
angle to the the accretion region.

The is some degree of uncertainty in the viewing angle to the
accretion regions. Tuohy et al (1988) found $i\sim40^\circ$ and the
red pole (the bright X-ray pole) has $\beta\sim125^\circ$ and the blue
pole (the faint X-ray pole), $\beta\sim40^\circ$. More detailed
modelling of Cropper et al (1990) and Wickramasinghe et al (1991) find
$i\sim40^\circ$, a dipole offset of 165$^\circ$ and the red pole
displaced from the magnetic pole by $\sim30^\circ$ and the blue pole
$\sim150^\circ$. However, the blue accretion region (the faint X-ray
region) is seen for all spin phases (and hence a relative small
viewing angle) and the red region for only a small fraction of the
spin phase (and hence at a high viewing angle). For argument we apply
a mean viewing angle of $\theta$=60$^\circ$ (implying sec$\theta$=2.0)
for the faint phase and $\theta$=80$^\circ$ (sec$\theta$=5.8) for the
bright phase.

We show in Table \ref{specfit} the resulting luminosities for the
shocked component (the `hard' component) and the blackbody component
assuming $kT_{bb}$=30eV. We find that the ratio, $L_{soft}/L_{hard}$,
in the faint phase is very low, with an upper limit of $\sim$0.1. In
the bright phase, the upper limit is 0.7.  In the standard accretion
model, hard X-rays in the post-shock region irradiate the photosphere
of the white dwarf and are re-emitted at lower energies. Assuming the
re-processed component is emitted in soft X-rays, we expect
$L_{soft}/L_{shock}\sim0.5$ (Lamb \& Masters 1979, King \& Lasota
1979). The total shock luminosity is the sum of hard X-ray emission
and the emission from the cyclotron component which originates from
the post-shock flow as well. Therefore, $L_{hard}$ underestimates
$L_{shock}$.

In the faint phase, it is clear that for a blackbody of temperature
30eV, the resulting energy balance is not consistent with the standard
shock model. However, lower temperature blackbodies can be `hidden' at
EUV wavelengths. To investigate this further, we added a blackbody
component of different temperatures and set their normalisation so
that the derived flux matched that observed in the UVW2 filter. We
show the model of the combined blackbody plus post-shock model in
Figure \ref{bb} for $kT_{bb}$=2, 5, 10 and 20eV. We also show in Table
\ref{bbtab} the fit to the data for these models in the 0.1--1.0keV
band, the resulting flux and luminosity for the unabsorbed
blackbody. We find that the observed X-ray spectra are not consistent
with blackbody temperatures greater than $\sim$10eV.

The un-heated surface of the white dwarf is also expected to
contribute to the flux seen in the UV. However, we do not know its
temperature (but it is expected to be near 1--2eV $\sim$10000-20000K)
nor do we have an accurate estimate of its distance. Because of these
uncertainties we do not add a blackbody to account for this component:
$\sim$10eV is therefore an approximate upper-limit to the temperature
of the reprocessed component. For $kT_{bb}$=10eV the resulting
$L_{soft}/L_{hard}$ ratio ($\sim$9) is very much greater than that
expected from the standard accretion shock model but for $kT_{bb}$=5eV
it is slightly higher than expected ($\sim$1.6). For lower
temperatures (2eV) the ratio is slightly lower than that expected.

\begin{table}
\begin{tabular}{llllr}
\hline
$kT_{bb}$ & \rchi & Blackbody flux & Blackbody & $L_{soft}/$\\
(eV) & (dof) & (\ergscm) & luminosity & $L_{hard}$\\
     &       & & (\ergsd) & \\
\hline
20 & 4.31 (27) & 2.68$\times10^{-10}$ & 1.6$\times10^{32}$& 64\\
10 & 1.39 (27) & 3.86$\times10^{-11}$ & 2.2$\times10^{31}$ & 9.0\\
5 & 1.43 (27) & 6.70$\times10^{-12}$ & 4.0$\times10^{30}$& 1.6\\
2 & 1.44 (27) & 1.34$\times10^{-12}$ & 8.0$\times10^{29}$& 0.3\\
\hline
\end{tabular}
\caption{Using the MOS spectra from the faint phase we have added a
blackbody of various temperatures: the normalisation has been set to
give the observed flux in the UVW2 filter. We show the resulting fit
to the X-ray data in the 0.1--1.0keV band, the unabsorbed bolometric
flux and luminosity of the blackbody component and the ratio
$L_{soft}/L_{hard}$ (cf Table \ref{specfit}).}
\label{bbtab}
\end{table}

\begin{figure*}
\begin{center}
\setlength{\unitlength}{1cm}
\begin{picture}(16,12)
\put(0,12.8){\includegraphics{faint_model_20ev_cap.ps}}
\put(8.,13.2){\includegraphics{faint_model_10ev_cap.ps}}
\put(-0.9,6.5){\includegraphics{faint_model_5ev_cap.ps}}
\put(7.7,6.6){\includegraphics{faint_model_2ev_cap.ps}}
\end{picture}
\end{center}
\caption{We show the best fit shock model that was used to fit the
EPIC MOS1 data in the faint phase together with a blackbody of various
temperatures. The normalisation of the blackbody was chosen to fit the
observed flux in the UVW2 filter. A reprocessed component with less than
$kT_{bb}\sim$10eV is therefore be cool enough not to be detected in
the X-ray band.}
\label{bb}
\end{figure*}

\subsection{The distance}

In the bright phase the hard X-ray bolometric luminosity is
$\sim1\times10^{31}$ \ergsd. To place a very crude estimate on the
distance to CE Gru we compare this luminosity to the hard X-ray
luminosity of AM Her (whose distance is reasonably well
determined). Ishida et al (1997) observed AM Her using {\sl ASCA} when
it was in a high accretion state and find $L_{hard}=1.6\times10^{32}$
\ergss using a distance of 75 pc. We find that CE Gru would have to
lie $\sim$400 pc distance to equal the $L_{hard}$ found for AM Her.

\section{Discussion}

Although CE Gru was detected in X-rays for the first time and was
clearly in a high accretion state, no distinct soft X-ray component
was observed. On the face of it, this is surprising since a strong
soft X-ray flux has long been considered one of the defining
properties of polars. Observations of polars using {\exo} (eg Osborne
1988) and {\ros} (eg Ramsay et al 1994) found a strong distinct soft
X-ray component, and in general, the ratio
$L_{soft}/L_{hard}\gg1$. Further, the ratio was correlated with the
magnetic field strength of the white dwarf: a high field gave a high
ratio. To account for the large `soft X-ray excess' seen in many
polars, dense `blobs' of material which do not form a shock and
radiate in soft X-rays were proposed (Kuijpers \& Pringle 1982, Frank,
King \& Lasota 1988).

Currently, four polars have been observed using {\xmm}. Of those
systems observed in an intermediate or high accretion state, WW Hor
showed no distinct soft X-ray component (Ramsay et al 2001) while BY
Cam has one pole which did show such a component and one which did not
(Ramsay \& Cropper 2002). Indeed only one other system, DP Leo, has
shown a distinct soft X-ray component (Ramsay et al 2001). Even for
those polars which showed the lowest ratios using {\ros} (EF Eri and
AM Her) a distinct soft X-ray component was still observed.

A variation of the standard shock model has been proposed by Heise \&
Verbunt (1988) and G\"{a}nsicke, Beuermann \& de Martino (1995) who
suggest that the reprocessed X-ray component lies in the EUV band
while the strong soft X-ray component seen in many polars originates
from dense `blobs' of material which do not form an exposed shock. Our
data are consistent with this view if the temperature of the
reprocessed component is $\sim2-5$eV: below 2eV the flux of the
reprocessed component is too small, while above 5eV it would be too
high. A component with a temperature above 10eV would be evident in
the spectrum (Figure~\ref{bb}). The implication in this case is that
the fraction of blobby accretion is small in CE Gru. Again this is
consistent with the unfolded light curves (not shown) which do not
show strong flaring seen in many systems (such as BY Cam -- Ramsay \&
Cropper 2002).
 
The amount of irradiation that the white dwarf receives is a function
of height in the post-shock region, with greater temperatures from
higher up, but more overall flux (even at higher energies) from near
the base (Cropper, Wu \& Ramsay 2000). On the other hand, because of
the curved surface of the white dwarf, the illumination is decreasing,
with slightly more than the square of the distance from the axis of
the post-shock region. Further, albedo varies as a function of
temperature. Although the effects of irradiating a white dwarf
atmosphere have been explored to some extent (Williams et al 1987,
Heise 1995), this is an area which needs further work to determine how
the temperature of the reprocessed spectrum is effected by parameters
such as the accretion rate and magnetic field.

We note, however, that higher specific accretion rates result in lower
shock heights, and higher magnetic fields also reduce the height of
the post-shock region (see for example Cropper et al 1999), so the
solid angle of emission from the post-shock flow intercepted by the
white dwarf photosphere is increased. In these high-state data of CE
Gru, we have no reason to expect anomalously low specific accretion
rates, and CE Gru has a magnetic field strength typical of
polars. Therefore there is no obvious reason for the reprocessed
component to have moved into the EUV if it is normally in the soft
X-ray band in polars. This gives support to the Heise \& Verbunt
(1988) suggestion that the soft X-ray component is indeed caused by
blobby accretion.

Some factor(s) must determine the number, length and density of blobs.
Obvious parameters include the magnetic field strength and orientation
of the white dwarf, the orbital period and the mass transfer
rate. Ramsay et al (1994) identified the magnetic field strength as
one important parameter.  Further progress in this regard awaits a
systematic analysis of the strength of the soft X-ray component as a
function of these parameters in a sufficiently large sample of polars.

\end{document}